\documentstyle[preprint,aps,version2]{revtex}
\tightenlines

\newcommand{\be}{\begin{equation}}
\newcommand{\ee}{\end{equation}}
\newcommand{\beq}{\begin{eqalignno}}
\newcommand{\eeq}{\end{eqalignno}}

\newcommand{\llike}{\ln {\cal L}}
\newcommand{\epem}{\mbox{$e^+e^-$}}
\newcommand{\gev}{{\rm\,GeV}}
\newcommand{\tev}{{\rm\,TeV}}

\newcommand{\ifb}{{\rm\,fb}^{-1}}
\newcommand{\mz}{M_Z}
\newcommand{\mw}{M_W}
\newcommand{\cosb}{\cos\beta}
\newcommand{\sinb}{\sin\beta}
\newcommand{\tanb}{\tan\beta}
\newcommand{\cosw}{\cos\theta_W}
\newcommand{\sinw}{\sin\theta_W}
\newcommand{\chc}{\tilde{\chi}^{\pm}}
\newcommand{\chcp}{\tilde{\chi}^+}
\newcommand{\chcm}{\tilde{\chi}^-}
\newcommand{\chn}{\tilde{\chi}^0}
\newcommand{\LSP}{\chn_1}
\newcommand{\sq}{\tilde{q}}
\newcommand{\msq}{m_{\tilde{q}}}
\newcommand{\msql}{m_{\tilde{q}_L}}
\newcommand{\msqr}{m_{\tilde{q}_R}}
\newcommand{\mmin}{m_{\tilde{q}}^{\rm min}}
\newcommand{\mprim}{m_{\tilde{\chi}}^{\rm primary}}
\newcommand{\mum}{(\mu , M_2)}
\newcommand{\noin}{\noindent}
\newcommand{\mpt}{\rlap{$\not$}p_T}
\newcommand{\sQl}{\tilde{Q}}
\newcommand{\sUr}{\tilde{U}}
\newcommand{\sDr}{\tilde{D}}
\newcommand{\su}{\tilde{u}}
\newcommand{\sd}{\tilde{d}}

\begin{document}
\draft

\pagestyle{empty}
\preprint{SLAC--PUB--6289}
\medskip
\preprint{hep-ph/9310211}
\medskip
\preprint{September 1993}
\medskip
\preprint{T/E}

\begin{title}
Squark Mass Determination at the \\
Next Generation of Linear \epem\ Colliders
\thanks{Work supported by the Department of Energy, contract
DE--AC03--76SF00515.}
\end{title}

\author{
Jonathan L. Feng
\thanks{Work supported in part by an NSF Graduate Research Fellowship.}
\thanks{E-mail address: {\tt jlf@slacvm.slac.stanford.edu}.}
and Donald E. Finnell
\thanks{E-mail address: {\tt dfinnell@slacvm.slac.stanford.edu}.}
}

\begin{instit}
Stanford Linear Accelerator Center\\
Stanford University, Stanford, California 94309
\end{instit}

\begin{abstract}

\centerline{ABSTRACT}
\bigskip

Current mass limits allow the possibility that squarks may be produced
in large numbers at the next generation of linear \epem\ colliders. In
this paper we investigate the prospects for precision studies of squark
masses at such colliders. We assume that squarks are lighter than
gluinos, and discuss both direct and cascade decay scenarios. By
exploiting the clean environment and polarizable beams of linear \epem\
colliders, we find that squark mass determinations at the level of a
few GeV are possible in a large part of the parameter space.

\end{abstract}

\centerline{(Submitted to Physical Review {\bf D})}

\newpage
\pagestyle{plain}

\narrowtext

\section{Introduction}
\label{sec:intro}

Supersymmetry (SUSY) is at present a promising theoretical framework
for physics beyond the Standard Model \cite{Reviews}. In addition to
being free of quadratic divergences and providing a natural dark matter
candidate, the simplest supersymmetric theory, the Minimal
Supersymmetric Standard Model (MSSM), has had remarkable success in
explaining the unification of coupling constants \cite{ccunif}.  To
solve the gauge hierarchy problem, SUSY must be broken at energies of
order 1 TeV, and thus the supersymmetric particles of the MSSM must be
within reach of the next generation of accelerators. This has
stimulated a great deal of activity in MSSM phenomenology in recent
years.  It should be noted, however, that most of this activity has
centered on particle searches. If supersymmetry is discovered, there
will be a rich spectrum of superparticles, and detailed studies of
their masses and couplings will be the focus of experimental particle
physics into the next century.  Precision measurements of such
quantities are crucial to the understanding of the SUSY breaking sector
of the MSSM and may even inform attempts to understand the mechanism of
SUSY breaking in supergravity and string theories.  It is not, then,
premature to investigate the prospects for detailed study of
superparticle properties at future facilities.

SUSY events are commonly characterized by unobservable particles in the
final state, and it is therefore not obvious that the underlying SUSY
parameters may be extracted from them with enough precision to be of
theoretical interest. To date there has been relatively little work in
this direction, with the exception of recent precision studies of
sleptons and neutralinos/charginos
\cite{JLC,Orito,Kon,Tsukamoto,Vandervelde}. These studies have shown
that if such particles are pair-produced at a future \epem\ collider,
their masses could indeed be determined with enough accuracy to be
significant tests of grand unified supergravity models. These particles
are a natural starting point, as they are expected to be the lightest
in the supersymmetric spectrum and therefore the most likely to be
accessible at \epem\ colliders with $\sqrt{s} = 500 \gev - 1 \tev$. It
is also possible, however, that squarks may be light enough to be
produced at such machines. This scenario is allowed by current
experimental limits \cite{CDF} and, if true, would provide an extremely
fertile ground for future experiments. Present limits allow squarks
with masses above 100 GeV.  With the expected features of the next
\epem\ collider, luminosities of $10 \ifb /{\rm year}$ and beam
energies of 250 GeV \cite{JLC,Ahn,Zerwas}, these would be pair-produced
at the rate of thousands per year. More massive squarks would have to
wait for accelerator upgrades, but their study would be qualitatively
similar. Although squark searches have been considered in great detail
in the context of hadron colliders
\cite{Gelmini,Gamberini,Drees,Woodside,Bartl1,Bartl2}, precision
studies at these machines would be very difficult.  Not only must one
control large hadronic backgrounds, but one must also work in a
situation where only a few features of a complex event are
reconstructed.  In contrast, the clean environment of \epem\
accelerators makes them promising for such studies. An added advantage
is the availability of polarized beams, which, as we will see, is a
very useful tool.  The purpose of this article is to survey the
prospects for precise determination of squark masses in the \epem\
environment.

The discussion is organized as follows:  In Sec.~\ref{sec:mssm} we
review relevant information about squarks and the MSSM. We will note
precisely which simplifying assumptions of the MSSM we use in our
analysis. In Sec.~\ref{sec:plane} we describe our event simulation and
organize our exploration of the SUSY parameter space. In
Sec.~\ref{sec:direct} we treat the simplest region, in which squarks
decay directly to the lightest supersymmetric particle (LSP).  The more
complicated regions with cascade decays are investigated in
Secs.~\ref{sec:gaugino} and \ref{sec:higgsino}, where the light
neutralinos are gaugino-like and higgsino-like, respectively. In
Sec.~\ref{sec:higher} we extend the analysis to the case of higher mass
squarks and higher energy colliders.  We conclude with some final
remarks in Sec.~\ref{sec:conc}.

\section{Squarks in the MSSM}
\label{sec:mssm}

The MSSM is the simplest extension of the Standard Model that
incorporates supersymmetry.  Here we concentrate on the salient
features for squarks and their decays.  Full discussions may be found
in a number of articles \cite{Reviews,GH}.

We will use hats to denote superfields and tildes to denote
superpartners of Standard Model particles.  The MSSM includes the usual
matter superfields and two Higgs doublet superfields

\begin{equation}\label{Hfields}
\hat{H}_1=\left( \begin{array}{c} \hat{H}_1^0 \\
                                  \hat{H}_1^- \end{array} \right)
\hspace{.2in} {\rm and} \hspace{.2in}
\hat{H}_2=\left( \begin{array}{c} \hat{H}_2^+ \\
                                  \hat{H}_2^0 \end{array} \right),
\end{equation}
where $\hat{H}_1$ and $\hat{H}_2$ give masses to the isospin
$-\frac{1}{2}$ and $+\frac{1}{2}$ fields, respectively.  These two
superfields are coupled in the superpotential through the term $ - \mu
\epsilon_{ij}\hat{H}_1^i\hat{H}_2^j$, where $\mu$ is the supersymmetric
Higgs mass parameter.  Both $\mu$ and $\tanb \equiv \langle H^0_2
\rangle / \langle H^0_1 \rangle$ will be important for our analysis of
squark decays. Soft supersymmetry breaking terms \cite{Girardello} for
scalars and gauginos are included in the MSSM with

\begin{equation}\label{vsoft}
V_{\rm soft} = \sum_i M_i^2 |\phi_i |^2 + \frac{1}{2} \left\{
[M_1\tilde{B}\tilde{B} + \sum_{j=1}^{3} M_2\tilde{W}^j\tilde{W}^j +
\sum_{k=1}^{8} M_3\tilde{g}^k \tilde{g}^k] + {\rm h.c.}\right\},
\end{equation}
where $i$ runs over all scalars.

For every flavor of quark in the Standard Model, the MSSM contains both
a left- and a right-handed squark whose masses are given by

\be\label{masses}
\begin{array}{lll}
m_{\su_L}^2 & = & M_{\sQl}^2+m_u^2 +
m_Z^2\cos{2\beta}\, (\frac{1}{2}-\frac{2}{3}{\sin^2}{\theta_W}) \\
m_{\su_R}^2 & = & M_{\sUr}^2+m_u^2 + \frac{2}{3} m_Z^2
\cos{2\beta}\, {\sin^2}{\theta_W}         \\
m_{\sd_L}^2 & = & M_{\sQl}^2+m_d^2 -
m_Z^2\cos{2\beta}\, (\frac{1}{2}-\frac{1}{3}{\sin^2}{\theta_W}) \\
m_{\sd_R}^2 & = & M_{\sDr}^2+m_d^2 -\frac{1}{3} m_Z^2
\cos{2\beta}\, {\sin^2}{\theta_W},
\end{array}
\ee
where we have suppressed generational indices and have neglected
left-right mixing terms which are relevant only for the top and bottom
squarks. The terms $M_{\sQl}$,$M_{\sUr}$, and $M_{\sDr}$ are some of
the soft supersymmetry breaking terms $M_i$ of Eq.~\ref{vsoft}, and the
remaining terms are model independent and known once $\tan\beta$ has
been determined.  Accurate determination of squark masses thus allows
determination of the SUSY breaking parameters of the squark sector,
which are of great theoretical interest.

In the most general form of the MSSM, the different SUSY breaking
parameters are unrelated.  To satisfy flavor changing neutral current
constraints, however, corresponding terms in the first and second
generations must be nearly degenerate \cite{FCNC}. (Recent work has
noted that adequate squark degeneracy may be enforced by gauged
horizontal symmetries \cite{Alex} or may in fact be unnecessary to
satisfy the FCNC constraints \cite{Nir}.) In minimal low energy
supergravity models, the SUSY breaking parameters are usually related
by assuming that they evolve from a universal scalar mass at a high
scale. Renormalization group evolution then predicts a splitting
between the left- and right-handed terms due to the contribution of
$SU(2)$ gaugino loops, which can be of order 5\% or higher depending on
the gaugino mass, as well as a splitting of the third generation
squarks from the first two. However, this assumption of a universal
scalar mass is not on firm theoretical footing, and one of the goals of
precision mass determinations is to differentiate between the universal
scalar mass scenario and one where the different masses are not so
simply related. We therefore do not assume a universal scalar mass.

The experimental signature of squark production will depend crucially
on the details of the squark decay channels. As we will see below, the
particles of the Higgs and gaugino sectors appear as intermediate decay
states, and it is therefore necessary to discuss these sectors of the
theory.  The Higgs sector consists of two CP-even scalars $h^0$ and
$H^0$, the charged scalar $H^\pm$, the CP-odd scalar $A^0$, and the
Higgsinos $\tilde{H}_1^-$, $\tilde{H}_2^+$, $\tilde{H}_1^0$, and
$\tilde{H}_2^0$.  The Higgsinos mix with gauginos to form mass
eigenstates.  The charged mass terms are $(\psi ^-)^T {\bf M}_{\chc}
\psi^+ + {\rm h.c.}$, where $(\psi^{\pm})^T = (-i\tilde{W}^{\pm},
\tilde{H}^{\pm})$ and

\be\label{chamass}
{\bf M}_{\chc} = \left( \begin{array}{cc}
 M_2                    &\sqrt{2} \, \mw\sinb  \\
\sqrt{2} \, \mw\cosb   &\mu                    \end{array} \right).
\ee
The chargino mass eigenstates are $\tilde{\chi}^+_i = {\bf
V}_{ij}\psi^+_j$ and $\tilde{\chi}^-_i = {\bf U}_{ij}\psi^-_j$, where
the unitary matrices {\bf U} and {\bf V} are chosen to diagonalize
${\bf M}_{\chc}$.  Neutral mass terms may be written as $\frac{1}{2}
(\psi ^0)^T {\bf M}_{\chn} \psi^0 + {\rm h.c.}$, where $(\psi^0)^T =
(-i\tilde{B},-i\tilde{W}^3, \tilde{H}^0_1, \tilde{H}^0_2)$ and

\be\label{neumass}
{\bf M}_{\chn} =
 \left( \begin{array}{cccc}
M_1             &0              &-\mz\cosb\sinw &\mz\sinb\sinw  \\
0               &M_2            &\mz\cosb\cosw  &-\mz\sinb\cosw \\
-\mz\cosb\sinw  &\mz\cosb\cosw  &0              &-\mu           \\
\mz\sinb\sinw   &-\mz\sinb\cosw &-\mu           &0     \end{array}
\right).
\ee
The neutralino mass eigenstates are $\chn_i = {\bf N}_{ij}\psi^0_j$,
where {\bf N} diagonalizes ${\bf M}_{\chn}$.  The four neutralinos in
increasing order of mass are $\chn_1$, $\chn_2$, $\chn_3$, and
$\chn_4$, and the two charginos, similarly ordered, are $\chc_1$ and
$\chc_2$.

The MSSM as reviewed above contains many arbitrary constants, and
usually some simplifications are made.  In our version of the MSSM we
will employ the following assumptions:

\noin (a) We assume R-parity conservation, a common assumption that
imposes baryon and lepton number conservation.  R-parity conservation
implies that the LSP is stable and must be among the decay products of
any supersymmetric particle.

\noin (b) We take the LSP to be the lightest neutralino, $\LSP$.  The
LSP must be neutral, and the other candidate, the sneutrino
$\tilde{\nu}$, is heavily disfavored if one assumes that the LSP makes
up galactic dark matter \cite{Caldwell}.  With this assumption, squarks
can either decay directly into an LSP and a quark,

\be\label{direct}
\sq\rightarrow q\LSP,
\ee
or indirectly to an LSP through a chain of neutralinos and charginos,
{\em e.g.},

\be\label{cascade}
\sq\rightarrow q'\tilde{\chi}^+_1\rightarrow q'\LSP (q\bar{q},
\nu\bar{l})         \hspace{.3in} {\rm or } \hspace{.3in}
\sq\rightarrow q \chn_2\rightarrow q \LSP (q\bar{q}, l\bar{l},
\nu\bar{\nu}).
\ee
The latter decays are called cascade decays.  Even more complicated
cascades involving $\chn_3$, $\chn_4$, and $\chc_2$ are possible. The
LSP interacts very weakly and disappears from the detector like a
neutrino, so direct LSP decays leave a distinctive signature of
acoplanar jets + missing $p_T (\mpt )$. In the case of cascade decays,
there may be additional jets and leptons.

\noin (c) We assume $\msq < |M_3|$ so that squarks do not decay through
gluinos. Without this assumption, the decay $\tilde{q} \rightarrow q
\tilde{g}$ would be possible and would in fact dominate.  Since the
gluino is strongly interacting, it would be accompanied by additional
hadronic radiation and a weakened $\mpt$ spectrum.  In such a case, a
different analysis from ours will be needed.

\noin (d) In this discussion, we assume that the neutralino/chargino
sector has already been explored.  We will therefore keep this sector
simple by assuming the unification of gauge constants and gaugino
masses at some higher scale. This implies that even at lower scales we
have the one-loop condition \cite{Inoue}

\begin{equation}\label{gaugunif}
\frac{M_2}{g^2_2} = \frac{3}{5} \frac{M_1}{g'^2} = \frac{M_3}{g^2_3}.
\end{equation}
Eq.~\ref{gaugunif}, coupled with assumption (c), will make our analysis
invalid in some areas of the $\mum$ plane, which will be noted below.

Because direct decays of the squarks to the LSP are characterized by
two acoplanar jets with large $\mpt$, they should be easy to resolve
experimentally. For example, cutting events with $\mpt < 35$ GeV and
$\theta_{acop} > 150^{\circ}$ will eliminate the bulk of Standard Model
backgrounds \cite{Grivaz} while eliminating only 20\% -- 40\% of the
signal events. The primary backgrounds after these cuts are $W^+W^-$
production, $e^{\pm}\nu W^{\mp}$ {\it via} $\gamma W$ fusion, and $\nu
\bar{\nu} Z$ {\it via} $WW$ fusion.  $W$ pair production is a
background to direct squark decays when one $W$ decays leptonically and
the charged lepton is either mistakenly included in a jet or goes
undetected.  In the former case one can eliminate events whose missing
momentum and energy are consistent with an undetected neutrino, and in
the latter case one can eliminate events whose visible mass is
consistent with the $W$ mass. All in all, we expect $W$ pair production
to be a less troublesome background for squark studies than it is for
chargino and slepton studies.  A more serious background is $e^{\pm}\nu
W^{\mp}$.  Before cuts, this cross-section is roughly an order of
magnitude above the signal, and because the electron tends to be lost
down the beam pipe and the neutrino tends to be produced with large
$p_T$, the cuts are not particularly effective. However, this
background can be removed by a cut on the two jet invariant mass. By
eliminating events with invariant mass less than 100 GeV the background
is effectively removed while much of the signal is retained. Such a cut
also reduces the $\nu \bar{\nu} Z$ background to negligible levels. One
might also be able to take advantage of the fact that these backgrounds
tend to produce jets in the same hemisphere, while squark pair jets are
preferentially in opposite hemispheres.

Cascade decays result in a wide variety of signals, and are therefore
generally more difficult to isolate.  However, in the analysis of this
paper, we will concentrate primarily on events with two jets and
isolated leptons.  Backgrounds to such events consist largely of events
where an on-shell $W$ or $Z$ decays hadronically, and these will
therefore also be removed by a cut on the two jet invariant mass.  One
important exception to this is $t \bar{t}$ events, which should be
well-understood.  This is a background when both top quarks decay to
$bl\nu$.  We can reduce this background by anti-tagging $b$ quarks, and
in our discussion of cascade decays, we will discuss the effectiveness
of such a cut.  This cut excludes $b$ and $t$ squarks from our
analysis: however, for reasons outlined in the following section, the
third generation of squarks will most likely require a separate
analysis anyway.

In addition to the Standard Model backgrounds discussed above, there
may also be SUSY backgrounds to consider, such as chargino and
neutralino production. Chargino pairs are a background to 2-jet cascade
events when one chargino decays through a  hadronic $W$ decay and the
other through a leptonic $W$ decay.  Neutralino production can result
in backgrounds to both direct and cascade squark decays.  However, in
both chargino and neutralino 2-jet events, the two jets are produced by
an on- or off-shell $W$ and $Z$ decay, so the two jet invariant mass
should be less than or of order the $W$ and $Z$ masses.  Thus, the
invariant mass cuts that reduce the Standard Model backgrounds should
also effectively reduce the SUSY backgrounds. In any case, it is not
unreasonable to hope that once the Standard Model and MSSM backgrounds
are well understood, efficient cuts can be devised to isolate the
squark signal.

In the remainder of this paper we will apply only the $\mpt$ and
acoplanarity cuts given above, namely $\mpt < 35$ GeV for all events
and $\theta_{acop} > 150^{\circ}$ for events with only two jets.  A cut
requiring the 2-jet invariant mass to be greater than 100 GeV has only
a small effect on the distributions we will plot, though it decreases
the total signal by roughly a factor of 3.

\section{Event Simulation and Organization of Parameter Space}
\label{sec:plane}

For the purposes of this exploratory study, we use a simple parton
level Monte Carlo to simulate squark production and decay.  We then
simulate hadronization and detector effects by smearing quark jets with
a detector resolution of $\sigma_E^{\rm had}/E = 50\%/\sqrt{E}$ ($E$ in
GeV). The cuts on $\mpt$ and $\theta_{acop}$ discussed above are
implemented to separate Standard Model backgrounds. Beamstrahlung and
initial state radiation are not included, and we assume 100\% electron
beam polarization.

Systematic errors in this study are of two kinds.  There are, of
course, systematic errors arising from hadronization and detector
effects. As our modeling of these effects is rather crude, a detailed
study of these errors will not be attempted here.  In addition, there
are errors arising from the uncertainty in the determination of the
neutralino and chargino masses which enter our analysis. The effects of
these errors on our ability to determine squark masses will be included
in our discussion of the case of direct decays.

We must fix the squark masses for the Monte-Carlo simulations. To
simplify the analysis, we assume that all left-handed squarks from the
first two generations are degenerate, as are the right-handed squarks,
and concentrate on the left-right splitting. This assumption may be
relaxed without major changes to the analysis, but will result in
greater complications in fitting the data. The third generation of
squarks will require a separate analysis, as the heavy top mass may not
allow direct decays of the stop to the LSP, and in addition there will
be an appreciable left-right mixing. In most situations, the third
generation of squarks can be separated with $b$ anti-tagging, so we
will consider only the first two generations from here on.

The present lower mass limit on squarks is $\sim$ 100 GeV from the CDF
experiment \cite{CDF} when the important effects of cascade decays have
been included \cite{Woodside}.  These limits assume $M_3<400\gev$ and
disappear for higher $M_3$.  Squarks will be pair-produced by $\epem
\rightarrow (\gamma,Z) \rightarrow \sq \bar{\sq}$.  At \epem\ colliders
with $\sqrt{s} = 500 \gev$, the production rate becomes substantial not
far below the kinematic limit of $\msq = 250 \gev$. For our studies, we
will choose $\msql = 220 \gev$ and $\msqr =210\gev$. These values are
significantly above the present mass bounds, but are also low enough to
give $\approx$ 2500 events per year, assuming two degenerate squark
generations and unpolarized beams at a luminosity of $10 \ifb /{\rm
year}$.  The results for $\msq = 400 \gev$ and $\sqrt{s} = 1 \tev$ are
qualitatively similar; we will address this case in
Sec.~\ref{sec:higher}.

In Fig.~\ref{fig:sigma}, we plot the cross sections for pair production
of $\sq_L\bar{\sq}_L$ and  $\sq_R\bar{\sq}_R$ from \epem\ annihilation,
assuming polarized $e^-$ beams. Note the slow rise of the cross section
near threshold, characteristic of scalar particle production, which
precludes the use of cross section measurements alone for precision
mass determinations. The polarization dependence of the cross section
is an important feature.  We see that $e^-_L$ beams produce
$\sq_L\bar{\sq}_L$ pairs 91\% of the time, and $e^-_R$ beams produce
$\sq_R\bar{\sq}_R$ pairs 91\% of the time.  Beam polarization is
therefore an extremely effective way of separating the left- and
right-handed squarks.

Since squarks are expected to be among the heaviest superparticles, one
cannot ignore the possibility of cascade decays. We should therefore
consider a representative variety of values of the neutralino/chargino
sector parameters rather than just the limiting cases. To do this, we
must organize our survey of the parameter space. Given the constraint
(\ref{gaugunif}), all chargino and neutralino masses are given as
functions of $\tanb$, $\mu$, and one of the gaugino masses, which we
will take to be $M_2$. Thus, for a given value of $\tanb$, squark
decays are determined by the parameters $\mum$, and our task reduces to
exploring squark decays in this parameter plane.

The discussion will be limited to the part of the plane with $0\leq
M_2<1\tev$ and $|\mu|<1\tev$.  The constraint $M_2\geq 0$ may be
imposed without loss of generality.
The upper bound on $M_2$ results from
the fact that $M_2$ is a SUSY breaking parameter, and for SUSY to
naturally explain the electroweak scale, $M_2$ cannot be too large.
Although $\mu$ is not a SUSY breaking parameter, a large value for
$\mu$ would re-introduce the fine-tuning problem. For most of this
study, we will fix $\tanb=2$. The analysis for higher $\tanb$ is not
much different and will be deferred to Sec.~\ref{sec:conc}.

With the parameters chosen above, we may now begin our study in earnest
by dividing the $\mum$ plane into regions with similar squark decay
channels. The decay patterns will, of course, also be influenced by the
Higgs and slepton masses.  Since these masses are unknown, however, the
boundaries separating regions with different squark decays through
Higgs and sleptons are not fixed.  For this reason, we first divide the
parameter space into regions ignoring the Higgs and slepton decays; we
will then consider the effects of varying the Higgs and slepton masses
within each region. The regions are shown in Fig.~\ref{fig:plane} for
$\msq = 220 \gev$.  We can ignore the hatched regions in the upper left
and right corners, because there the squarks are lighter than $\LSP$,
which violates our assumptions. The cross-hatched region along the
$M_2=0$ and $\mu=0$ axes is experimentally ruled out by lower bounds on
superparticle masses \cite{PDG,LEP}.  In region 4, the squarks can
decay to four or more of the neutralinos/charginos. This results in
very complicated cascade patterns.  Though our methods may, in
principle, be extended to this region, we will not consider it further
in this paper. In what remains of the plane, the situation is greatly
simplified, because the three neutralinos/charginos to which the squark
can decay are always $\LSP$, $\chn_2$, and $\chc_1$. The only two-body
decays kinematically allowed in region 1 are those directly to the LSP.
In regions 2 and 3, squarks may also decay to either $\chn_2$ or
$\chc_1$ or both.  In region 2, $\LSP$, $\chn_2$, and $\chc_1$ are all
dominated by their gaugino components. This region is further
subdivided into region 2a, where $\chn_2$ and $\chc_1$ may decay to the
LSP through on-shell $W$ and $Z$ bosons, and region 2b, where decays
through on-shell $W$ and $Z$ decays are not possible. In region 3,
$\LSP$, $\chn_2$, and $\chc_1$ are all Higgsino-like, and again only
decays through off-shell $W$s and $Z$s are possible. Finally, each
region has a mirror region in the $\mu>0$ part of the plane.  These
mirror regions usually have the same decay patterns as their $\mu<0$
counterparts, so where this is true, we will analyze only the $\mu<0$
regions.  The only exception is region 3, and we will note the
different behavior for $\mu>0$ at the end of Sec.~\ref{sec:higgsino}.
In Fig.~\ref{fig:plane} the representative points in each region that
we will consider in detail in the following sections are marked.

In what follows, it will often be helpful to keep in mind the following
approximate relationships \cite{GH2}.  At $M_Z$, $M_1\approx
\frac{1}{2} M_2$ and $M_3\approx \frac{10}{3} M_2$.  For $|\mu|, M_2
\gg M_Z$,

\be\label{thumb}
\begin{array}{llllll}
m_{\chn_1} & \approx & \min\{|\mu|,\frac{1}{2} M_2\}, \hspace{.5in} &
m_{\chn_2} & \approx & \min\{|\mu|,M_2\},                      \\
m_{\chn_3} & \approx & \max\{|\mu|,\frac{1}{2} M_2\},               &
m_{\chn_4} & \approx & \max\{|\mu|,M_2\},                      \\
m_{\chc_1} & \approx & \min\{|\mu|,M_2\},                 &
m_{\chc_2} & \approx & \max\{|\mu|,M_2\}.
\end{array}
\ee
Note that in this approximation, $\chn_2$ and $\chc_1$ are virtually
degenerate throughout the plane. We will take advantage of this fact in
our analysis of cascade decays in Sec.~\ref{sec:gaugino}.

It should also be noted that our assumption that the squarks are
lighter than gluinos, coupled with the unification assumption of
Eq.~\ref{gaugunif}, implies that our analysis is only valid for
$M_2\agt 67 \gev$ (above the dotted line in Fig.~\ref{fig:plane}).
Also, if one makes the further assumption that one can compute the
squark mass by applying the renormalization group equations with the
desert hypothesis, one obtains

\be\label{rge}
m_{\tilde{q}}^2 \approx m_0^2 + 7 m_{1/2}^2,
\ee
where $m_0$ and $m_{1/2}$ are the squark mass and the gaugino mass at
the unification scale $M_U$, respectively. Since $m_0^2 > 0$ and
$m_{1/2}\approx 0.8 M_2$, this leads to the constraint $M_2\alt 0.5
\msq$, which in the present case implies that $M_2 \alt 110 \gev$.

\section{Direct Decays}
\label{sec:direct}

In region 1 only direct decays of the squarks to the LSP are allowed.
For the sake of concreteness, we will perform our Monte-Carlo
simulations at the representative point $\mum = (-500 \gev, 300 \gev)$.
The quark jet from the decay of the scalar particle $\sq \rightarrow q
\LSP$ will have a flat energy distribution with endpoints

\be\label{endpts}
E_{\rm max,min}=\frac{E_b}{2}
\left(1 \pm \sqrt{1- \frac{m_{\tilde{q}}^2}{E_b^2}}\,\right)
\left(1-\frac{M_{\LSP}^2}{m_{\tilde{q}}^2}\right),
\ee
where $E_b$ is the beam energy, and we have neglected the quark mass.
Thus, in theory, squark masses can simply be deduced from the
distribution's endpoints. Unfortunately, the simple flat shape will be
changed by cuts, and finite detector resolution and hadronization will
smear the endpoints. Of the 1764 (975) squark pair events produced by
the $e^-_L$ ($e^-_R$) beam, 1294 (683) survive the cuts.  The energy
distribution of the individual jets from the surviving events is given
in Fig.~\ref{fig:jetenergy1}, where detector resolution effects have
been included. We see that the location of the endpoints is rather
ambiguous.  One way to extract the squark mass is to apply a binned
likelihood fit to the jet energy distribution, with the logarithm of
the likelihood given by

\be\label{loglike}
\llike(\msql,\msqr) = \sum_{i} A_i(\msql,\msqr) \ln B_i - B_i,
\ee
where the sum is over all bins, $A_i(\msql,\msqr)$ is the expected
number of events in bin $i$ given hypothetical squark masses $\msql$
and $\msqr$, and $B_i$ is the measured number of events in bin $i$. The
actual values of $\msql$ and $\msqr$ are determined by maximizing
$\llike$, and the statistical error of the determination is given by
the width of the $\llike$ peak. For simplicity we leave $\msqr$ fixed
at its actual value and calculate $\llike (\msql)$. In the $\llike$
calculations, we have approximated the theoretically expected number of
events $A_i$ by Monte-Carlo simulations with a very large number of
events (typically 50,000). We find that the $\msql$ determination has a
statistical error of 0.9 GeV at 95\% CL. These calculations have been
performed assuming that the LSP mass is known. The expected statistical
errors on the masses of $\chc_1$ and $\LSP$ from chargino studies are
3.2 and 2.0 GeV, respectively, while slepton studies should give the
LSP mass to 1.0 GeV \cite{JLC}. A shift of 1.0 GeV in the LSP mass
causes a shift of about 0.7 GeV in the central value of the likelihood
fit.

In the single jet energy spectrum used above, correlations between the
energies and directions of the two jets of a given event are ignored,
since each point in the distribution represents only one jet. It is
thus worth thinking about whether there is a more efficient way to use
the information contained in the event sample. One possibility is to
retain the jet correlation information by using a two-dimensional
binning, but this necessitates a large jump in computing time to obtain
the $A_i$ distribution accurately.

Instead, the method we will use extensively is the following: For each
event, we calculate the quantity $\mmin$, the minimum squark mass
kinematically possible, given the two observed quark jet momenta. For
each event, we then get one value of $\mmin$, and we apply a binned
likelihood method to the $\mmin$ distribution. The quantity $\mmin$ is
easily determined (see Fig.~\ref{fig:circle}). We label the particle
momenta as $\sq (p_1)\rightarrow \LSP (p_3) q(p_4)$ and $\sq
(p_2)\rightarrow \LSP (p_5) q(p_6)$. The total visible momentum is then
$p_V = p_4 + p_6$. Because we know $E_b$ and have measured $E_1$ and
$E_2$, we can determine the LSP energies $E_3$ and $E_5$.  However, we
also know the LSP mass, so we can find the magnitudes $|\vec{p}_3|$ and
$|\vec{p}_5|$, and since $p_3+p_5=-p_V$, the vectors $\vec{p}_3$ and
$\vec{p}_5$ are constrained to lie on a circle $C$.  We then can
calculate the angles $\gamma$ and $\delta$ shown in
Fig.~\ref{fig:circle}. The minimum squark mass corresponds to the
maximum possible $|\vec{p}_2|$, and is given by

\be\label{sqminmax}
(\mmin )^2 = E_b^2-|\vec{p}_3|^2- |\vec{p}_4|^2 +
2 |\vec{p}_3||\vec{p}_4| (\cos\gamma\cos\delta -
\sin\gamma\sin\delta).
\ee
The distributions of $\mmin$ for left- and right-polarized electron
beams are sharply peaked at the underlying masses $\msql$ and $\msqr$,
respectively (see Fig.~\ref{fig:msqmin}). As before, for simplicity we
leave $\msqr$ fixed at its actual value and calculate $\llike (\msql)$.
We find that the statistical error of the $\msql$ determination is
reduced to 0.4 GeV at 95\% CL. Thus, we see that the $\mmin$ method
improves our statistical error significantly. If the cut $m_{\rm 2-jet}
> 100 \gev$ is necessary to reduce background, our event sample is
reduced by a factor of 3, but the statistical error increases only
slightly to 0.5 GeV at 95\% CL. We note, however, that a shift of 1 GeV
in the LSP mass now shifts the central value of the likelihood fit by
approximately 1.5 GeV. The $\mmin$ calculation is more strongly
dependent on the LSP mass, and therefore places a higher premium on its
accurate determination.

The greater power of the $\mmin$ distribution is manifested in its
sharp peak, because even slight variations in $\msql$ and $\msqr$ move
these peaks enough to create large differences between $A_i$ and $B_i$
in some bins. This implies that $\llike$ falls rapidly from its
maximum. There are a number of reasons for the sharp peak of the
$\mmin$ distribution.  Here we just note that, roughly speaking,
momentum vectors lying on large circles $C$ may give mass minima both
close and far from the actual squark mass, depending on where the
momentum vectors lie on $C$. However, small circles give only accurate
solutions, and thus the calculated minimum masses preferentially lie
close to the actual underlying squark mass.

One potentially powerful feature of squark mass studies in this region
is that, since both left- and right-handed squarks have identical decay
channels, a direct comparison can be made to determine left-right mass
splittings.  The left- and right-handed squarks can be isolated using
polarized beams, and systematic errors, which should effect both
polarizations equally, should largely cancel in the ratio of their
masses.  One can therefore determine left-right mass splittings to
greater accuracy than one can determine the actual values of the
masses.  This is in contrast to the case of slepton studies, where use
of the left-polarized beam is hampered by a large $W^+W^-$ background
\cite{Orito,Vandervelde}. As was explained above in
Sec.~\ref{sec:mssm}, we do not expect $W^+W^-$ backgrounds to be a
signifigant problem for squarks.

While it may come as no surprise that precise mass determinations can
be made in the simple case of direct decays, the large squark masses
expected make it likely that more complicated decays will be present.
For a generic point in parameter space, with a large number of possible
decay chains present simultaneously, it is important to determine
whether it is still be possible to extract accurate squark masses from
the more complicated signal.  We now turn our attention to this
question in representive regions of parameter space.

\section{Gaugino-like Cascades}
\label{sec:gaugino}

In region 2 squarks have new decay channels through on-shell $\chn_2$
and $\chc_1$.  In this region, however, $\LSP\approx \tilde{B}$,
$\chn_2 \approx \tilde{W}^3$, and $\chc_1 \approx \tilde{W}^{\pm}$, so
since right-handed squarks do not couple to $SU(2)$ gauginos, they
still decay predominantly directly to the LSP.  This may be seen from
the branching ratios of $\tilde{u}_L$ and $\tilde{u}_R$, given in the
contour plots of Fig.~\ref{fig:br}.  Similar plots for down-type
squarks differ little. The analysis for right-handed squarks is
therefore simple. First we use the polarized $e^-_R$ beam to isolate
right-handed squarks. Contamination from left-handed squarks will be of
order 10\%, and most of these will go through complicated decay
channels and can be easily separated by considering only two-jet
events. We then apply the analysis of region 1 with little degradation
of statistics. As shown in Fig.~\ref{fig:br}, left-handed squarks,
unlike right-handed squarks, do decay predominantly through cascades in
some parts of region 2. When this is the case, a separate analysis for
left-handed squarks is necessary. The rest of this section will be
concerned with left-handed squarks only.

As suggested at the end of Sec.~\ref{sec:plane}, we can use the
near-degeneracy of $\chn_2$ and $\chc_1$ to simplify matters.
Throughout region 2, these two particles are typically degenerate to a
few GeV, and since

\be\label{massdiff}
|m_{\chn_2}-m_{\chc_1}| \ll \frac{1}{2} M_2 \approx m_{\chn_2}-m_{\LSP},
\ee
phase space suppression allows us to safely ignore decays of $\chn_2$
to $\chc_1$ and vice versa.  We are then left with two-step decays with
squarks decaying to $\chn_2$ and $\chc_1$, which then decay to the LSP
through three-body modes mediated by $W$ or $Z$ bosons, sleptons,
sneutrinos, squarks, or Higgs bosons.  The resulting quark jet energy
distribution is much more complicated than it was in the direct case,
with quarks produced both at the primary vertices, {\it i.e.}, the
initial squark decay vertices, and in the later cascade decays.

However, by choosing appropriate cuts, we can reduce the problem to the
case of direct decays. Recall that in the $\mmin$ analysis of region 1,
we needed to know only the energies and momenta of the quarks leaving
the primary vertices and the mass of the neutralino or chargino leaving
those vertices, which we now denote $\mprim$.  In the case of region 1,
such identifications are obvious, since in all events both of the quark
jets are produced at the primary (and only) vertices, and $\mprim =
m_{\LSP}$ is always the correct choice.  Cascade decays complicate this
analysis, because the initial squark decay may involve
neutralinos/charginos other than the LSP, and there may be many quarks
produced, making it difficult to determine which two came from the
primary vertices. Our stategy will be to find cuts that allow us to
accurately identify the primary quarks and assign $\mprim$. We can then
calculate $\mmin$ for each event and proceed as in the previous
section. Two simple strategies are (1) separating direct and cascade
decays kinematically and choosing events that contain direct decays on
both sides, and (2) considering double cascade events in which the
$\chn_2$ or $\chc_1$ has only leptonic decay products.  We will show
that these two strategies are always sufficiently effective. We now
consider regions 2a and 2b in turn.

Decays of $\chn_2$ and $\chc_1$ through on-shell $W$s and $Z$s are
allowed in region 2a.  We will take the representative point to be
$\mum = (-500 \gev, 200 \gev)$. At this point, the masses of $\LSP$,
$\chn_2$, and $\chc_1$ are 103.2, 206.1, and 206.0 GeV, respectively.
The jet energy distribution is shown in Fig.~\ref{fig:jetenergy2a},
where every quark jet produced in the Monte-Carlo squark decays is
binned separately. The distribution consists of three parts --- a large
peak of soft quarks emanating from squarks decaying to $\chn_2$ and
$\chc_1$ (long dashes), a flat distribution from direct LSP decays
(solid), and a wide hump from hadronic $W$ and $Z$ decays (short
dashes). The total distribution is given by the dotted histogram. The
soft jets in the peak may not be discernible experimentally, but our
analysis will use only jets from direct decays, and these jets have
energies that are always greater that 40 GeV.

As noted previously, since region 2 allows cascade decays, top quark
production is potentially a troublesome background.  In
Fig.~\ref{fig:bquarks}, we again plot the total jet energy distribution
resulting from squark decays, but this time, for comparison, also the
bottom quark spectrum resulting from top decays.  In the bottom quark
spectrum, we have removed all $b$ quarks produced in top events in
which at least one $b$ quark has been successfully tagged. We assume
$m_{\rm top} = 150 \gev$ and a $b$-tagging efficiency of 80\%
\cite{JLC}. The top quark background is substantially reduced below the
signal, and as top quark decays will be well-understood by the time of
these squark studies, the top quark background should not be a
significant obstacle.

The branching ratios at $\mum = (-500 \gev, 200 \gev)$ are

\be\label{br}
\begin{array}{lcrlcr}
{\rm BR}(\tilde{u}_L\rightarrow u\LSP)&=&58\%, & \hspace{.2in}
{\rm BR}(\tilde{d}_L\rightarrow d\LSP)&=&36\%,  \\
{\rm BR}(\tilde{u}_L\rightarrow u\chn_2)&=&28\%, & \hspace{.2in}
{\rm BR}(\tilde{d}_L\rightarrow d\chn_2)&=&43\%,  \\
{\rm BR}(\tilde{u}_L\rightarrow d\chcp_1)&=&14\%, & \hspace{.2in}
{\rm BR}(\tilde{d}_L\rightarrow u\chcm_1)&=&21\%,
\end{array}
\ee
so we see that there are still many direct LSP decays. We will
therefore try to isolate the double direct LSP decay events and then
apply the analysis of Sec.~\ref{sec:direct}.  Of the 1764 events
produced with a left-polarized beam in 1/2 year, 1437 pass the $\mpt$
and $\theta_{acop}$ cuts.  We then consider only events with 2 jets and
no isolated leptons, leaving 673 events, and for each of these we
calculate $\mmin$ using $\mprim = m_{\LSP}$. Of the remaining events,
244 are actually cascade events with neutrino decay products, but these
may be removed by considering only events in which both jets have
energy above 30 GeV. We find that the likelihood fit to the remaining
$\mmin$ distribution gives squark masses to 1.3 GeV at 95\% CL.

The distribution of Fig.~\ref{fig:jetenergy2a} was calculated for the
case where the sleptons and Higgs scalars are massive enough that their
diagrams are off-shell and suppressed.  Smaller slepton and Higgs
masses will, of course, change the part of the jet distribution
resulting from cascades, but will have no effect on the two-jet event
sample and will therefore not change our results.

We now turn to region 2b, whose representative point we take to be
$\mum = (-500 \gev, 100 \gev)$. Here the masses of $\LSP$, $\chn_2$,
and $\chc_1$ are 52.8, 108.1, and 107.8 GeV, respectively, and we see
that the $W$ and $Z$ diagrams are indeed off-shell.  The branching
ratios are

\be\label{br2}
\begin{array}{lcrlcr}
{\rm BR}(\tilde{u}_L\rightarrow u\LSP)&=&7\%, & \hspace{.2in}
{\rm BR}(\tilde{d}_L\rightarrow d\LSP)&=&1\%,  \\
{\rm BR}(\tilde{u}_L\rightarrow u\chn_2)&=&61\%, & \hspace{.2in}
{\rm BR}(\tilde{d}_L\rightarrow d\chn_2)&=&67\%,  \\
{\rm BR}(\tilde{u}_L\rightarrow d\chcp_1)&=&32\%, & \hspace{.2in}
{\rm BR}(\tilde{d}_L\rightarrow u\chcm_1)&=&32\%.
\end{array}
\ee
Because there are not many direct LSP decays, the analysis must rely on
cascade events.  This complicates the analysis.  However, we now show
that even for the most difficult sets of parameters, a significant
number of $\chn_2$ and $\chc_1$ decays will be purely leptonic.  In
these events we may unambiguously identify the two primary vertex quark
jets, and we can then again apply the region 1 analysis.

To do this we must analyze the relative importance of the various
cascade diagrams (see Fig.~\ref{fig:feyn}). The Higgs diagrams may be
safely ignored. Typical values for the mass of the lightest Higgs
scalar, $h^0$, are in the range 70--110 GeV. For this entire range, the
$h^0$ process proceeds off-shell and is also suppressed by the bottom
quark Yukawa coupling. The other Higgs scalars are significantly more
massive and can also be safely ignored. In addition, we have squark and
slepton cascade diagrams. Sleptons are generally expected to be lighter
than squarks, and we will see below that the lower the slepton masses,
the simpler our analysis. For now, we will pessimistically take the
sleptons to be degenerate and of mass 200 GeV.

Since we have squark and slepton masses significantly higher than
$\mw$, one might expect the $W$ and $Z$ diagrams to dominate.  However,
in the gaugino-like region of parameter space, $|{\bf N}_{11}| \approx
|{\bf V}_{11}| \approx 1$, so the LSP is primarily composed of
$\tilde{B}$, which does not couple to the $W$ or $Z$ at all.  Thus,
these diagrams are suppressed by a factor $S \equiv {\cal O}(|{\bf
V}_{12}|, |{\bf N}_{12}|, |{\bf N}_{13}|, |{\bf N}_{14}|)$.  We can
obtain a rough estimate of $S$ by taking an appropriate limit of the
explicit expression for ${\bf V}_{12}$ \cite{GH2}. Taking $\mw$ small
compared to $\mu$ and $M_2$, we have

\be\label{v12}
{\bf V}_{12} \approx \sqrt{2} \mw \frac{\mu\,\cosb + M_2\,
\sinb}{M_2^2-\mu^2} \approx
- \frac{50 \gev}{\mu} \left(1+\frac{2M_2}{\mu}\right),
\ee
for $\tanb=2$, where we have expanded in terms of $M_2/\mu$, a small
parameter in the gaugino-like region. Thus, ${\bf V}_{12}$ is about
0.02 at our representative point.  The ${\bf N}_{1j}$ are roughly of
the same order, and we find that $0.01 \alt S\alt 0.1$ throughout most
of the region. The competing suppressions of the different diagrams
make it impossible to simply determine which diagram dominates.  In
fact, our calculations show the gauge boson, squark, and slepton
diagrams to be roughly of the same order. (Note that if one takes the
gaugino-like relations $\LSP\approx -i\tilde{B}$, $\chn_2 \approx
-i\tilde{W}^3$, and $\chc_1\approx -i\tilde{W}^{\pm}$ too seriously,
one is led to incorrectly conclude that the $W$ and $Z$ diagrams may be
set to zero. This is never valid in the part of the plane we are
considering.)

Evaluating the various diagrams with $m_{\tilde{l}} = m_{\tilde{\nu}} =
200 \gev$, we find

\be\label{br3}
\begin{array}{lcrlcr}
{\rm BR}(\chn_2\rightarrow q\bar{q}\LSP)&=&20\%, & \hspace{.2in}
{\rm BR}(\chc_1\rightarrow q'\bar{q}\LSP)&=&55\%,  \\
{\rm BR}(\chn_2\rightarrow l\bar{l}\LSP)&=&31\%, & \hspace{.2in}
{\rm BR}(\chc_1\rightarrow l\nu\LSP)&=&45\%,  \\
{\rm BR}(\chn_2\rightarrow \nu\bar{\nu}\LSP)&=&49\%. & \hspace{.2in}
\end{array}
\ee
We see that there are many leptonic decays, and we can therefore
consider only the events with two quark jets + leptons without a great
loss in statistics. (Note that at this point in the parameter plane the
only jets below 40 GeV are those produced by $W$s and $Z$s.  Thus by
selecting only two-jet events, the jets that enter our analysis are
again all hard enough to be experimentally detected.) Of the 1508
$e^-_L$ polarized events that pass the $\mpt$ and $\theta_{acop}$ cuts,
834 have exactly 2 quark jets (+ leptons). Because the polarized beam
has eliminated most right-handed squarks from our sample, and
left-handed squarks rarely decay directly to the LSP, we expect that
most of these events are double cascade events, and in fact 70\% are.
We then calculate $\mmin$ with the assignment $\mprim = m_{\chc_1}$.
Note that since $\chn_2$ and $\chc_1$ are virtually degenerate, we need
not distinguish them for the kinematic analysis of
Sec.~\ref{sec:direct}. About half of the events with one or two direct
LSP decays are kinematically incompatible with the assignment $\mprim =
m_{\chc_1}$, and we eliminate these.  Almost none of the double cascade
events are removed by this cut. We are then left with 689 events, of
which 84\% are double cascade events. The double cascade events form a
sharp peak in the $\mmin$ distribution, while the few events involving
direct decays are much more broadly distributed. After performing a
likelihood fit to the $\mmin$ distribution, we find that the squark
mass can be determined to 2.4 GeV at 95\% CL.

If we lower the slepton mass, the slepton diagram contribution grows.
For $m_{\tilde{l}}=110 \gev$, even though the sleptons are still
off-shell, the gauge boson suppression factor $S$ defined earlier
allows the slepton diagrams to dominate, and the resulting branching
ratios are

\be\label{br4}
\begin{array}{lcrlcr}
{\rm BR}(\chn_2\rightarrow q\bar{q}\LSP)&=&0.3\%, & \hspace{.2in}
{\rm BR}(\chc_1\rightarrow q'\bar{q}\LSP)&=&6\%,  \\
{\rm BR}(\chn_2\rightarrow l\bar{l}\LSP)&=&39.8\%, & \hspace{.2in}
{\rm BR}(\chc_1\rightarrow l\nu\LSP)&=&94\%,  \\
{\rm BR}(\chn_2\rightarrow \nu\bar{\nu}\LSP)&=&59.9\%. & \hspace{.2in}
\end{array}
\ee
Clearly most $\chn_2$ and $\chc_1$ decay to leptons, and we can use
almost all of the cascade events in our analysis.

\section{Higgsino-like Cascades}
\label{sec:higgsino}

In region 3, $\chn_2$ and $\chc_1$ are again close in mass, but now, as
suggested in Eq.~\ref{thumb}, they are also close to the LSP in mass.
We will take as our representative point $\mum = (-100 \gev, 700
\gev)$, where the masses of $\LSP$, $\chn_2$, and $\chc_1$ are 98.3,
110.9, and 106.1 GeV, respectively.  The branching ratios of left- and
right-handed squarks to the LSP are now both $\sim$ 10\% -- 20\%. The
analysis for left- and right-handed squarks is similar, and we will
consider only the left-handed below.

The quark jet energy distribution is shown in
Fig.~\ref{fig:jetenergy3}. The large low energy hump is composed of the
soft quarks produced in the decays between the neutralinos and
chargino.  The primary vertex quarks have the flat energy distribution
we expect, but this is really a superposition of decays to three
different particles with slightly different masses.  We see that the
primary and secondary vertex quark jets are well-separated in energy,
and we can isolate the primary quark jets with a simple jet energy cut
at 30 GeV.  In fact, we again have not included the decays of $\chn_2$
to $\chc_1$ in the Monte-Carlo simulation. This omission is physically
unwarranted, as such decays no longer suffer the pronounced phase space
suppression relative to the decays of $\chn_2$ and $\chc_1$ to the LSP.
However, these decays will only increase the number of soft jets in the
low energy hump, which will be eliminated with the energy cut anyway.
We have also assumed $m_{\tilde{l}} = 200 \gev$ and $m_{h^0} = 110
\gev$.  Other values will change the shape of the low energy hump, but
again this is irrelevant after the energy cut.

We must now determine the squark mass from the jet energies.  Unlike in
the previous sections, we cannot use the $\mmin$ distribution, because
for a given quark jet, we cannot tell if its primary vertex partner was
$\chn_2$ or $\chc_1$, and the 5 GeV mass difference between these two
is now significant compared to the the accuracy with which we hope to
measure the squark masses. We will therefore simply use the jet energy
distribution for our likelihood fit. Our strategy will then rely on the
assumption that we know the neutralino and chargino parameters. The
$\chn$ and $\chc$ masses will be quite accurately determined and will
fix the relative positions of the endpoints of the energy
distributions.  Slepton and chargino studies should be able to
determine the parameters $\mu$ and $M_2$, which will determine the
branching ratios of the squarks. In the Higgsino-like region, it may be
difficult to determine $M_2$ precisely, but since the branching ratios
are relatively insensitive to $M_2$, we nevertheless also expect to
determine the branching ratios accurately. Given these assumptions, we
take the neutralino/chargino masses and branching ratios as inputs in
our analysis. The only unknown is then the squark mass, and a $\llike$
fit gives us the squark mass to within 1.2 GeV at 95\% CL.

Region 3 is the only region where the corresponding $\mu > 0$ region
has different decay properties.  In the $\mu > 0$ region, all types of
squarks decay to the LSP with branching ratios $\agt 80 \%$.  We can
therefore use the direct decays in this region, and the analysis is
actually simpler.

\section{Higher Squark Masses}
\label{sec:higher}

So far we have been studying squarks with mass $\sim 220 \gev$.  Of
course, squarks may be significantly more massive than this, and it is
therefore important to consider the applicability of the preceding
analysis to the case of higher mass squarks and higher energy
colliders.  With slight modifications, we will see that it is
straightforward to adapt the analysis to higher energies. We note,
however, that as the squark mass rises, our assumption of $\msq < M_3$
becomes more and more disfavored if one subscribes to the theoretical
prejudice of a universal scalar mass.

To proceed, we will consider the case of a 1 TeV collider with
luminosity $30 \ifb /{\rm year}$. Squarks with mass near the kinematic
limit of 500 GeV can be studied; we will take $\msql= 400 \gev$ and
$\msqr=390 \gev$. The $\mum$ plane may be divided into regions as
before (see Fig.~\ref{fig:1tevplane}).  All of the region boundaries
move to higher $|\mu|$ and $M_2$, except the boundary between regions
2a and 2b. The boundary at which gluinos become less massive than
squarks moves up to $M_2 \approx 121 \gev$. In addition,
renormalization group equatons coupled with the desert hypothesis now
imply $M_2 \alt 200 \gev$. Of course, region 4, the region of
complicated cascade decays, becomes large.  A full treatment of this
region would demand a significantly more complicated analysis.

The squark decays of regions 1, 2b, and 3 are qualitatively similar to
those in the lower energy case we considered earlier, and therefore
similar analyses are applicable. An important difference for our
analysis, however, is that region 2a is no longer a thin strip, and
consequently the branching ratio to the LSP is small throughout most of
region 2a.  In the previous discussion, the branching ratio to the LSP,
even for left-handed squarks, was substantial in region 2a, and we
based our analysis on the abundance of direct LSP decay. We see that
this convenient feature does not persist to cases of higher squark
masses and higher energy colliders, and for the case of 400 GeV squarks
we must use cascade decays for left-handed squarks in region 2a.

To study this in detail, we again choose the point $\mum = (-500 \gev,
200 \gev)$. Let us consider first the scenario when $m_{\tilde{l}}=380
\gev$ and $m_{h^0} = 110 \gev$, so sleptons and Higgs bosons are too
massive to be on-shell.  In this case, the branching ratios are

\be\label{brtev}
\begin{array}{lcrlcr}
{\rm BR}(\tilde{u}_L\rightarrow u\LSP)&=&5\%, & \hspace{.2in}
{\rm BR}(\tilde{d}_L\rightarrow d\LSP)&=&2\%,  \\
{\rm BR}(\tilde{u}_L\rightarrow u\chn_2)&=&63\%, & \hspace{.2in}
{\rm BR}(\tilde{d}_L\rightarrow d\chn_2)&=&66\%,  \\
{\rm BR}(\tilde{u}_L\rightarrow d\chcp_1)&=&32\%, & \hspace{.2in}
{\rm BR}(\tilde{d}_L\rightarrow u\chcm_1)&=&32\%,
\end{array}
\ee
and

\be\label{brtev2}
\begin{array}{lcrlcr}
{\rm BR}(\chn_2\rightarrow q\bar{q} \LSP)&=&68\%, & \hspace{.2in}
{\rm BR}(\chc_1\rightarrow q'\bar{q} \LSP)&=&66\%,  \\
{\rm BR}(\chn_2\rightarrow l\bar{l} \LSP)&=&11\%, & \hspace{.2in}
{\rm BR}(\chc_1\rightarrow l\nu \LSP)&=&34\%,  \\
{\rm BR}(\chn_2\rightarrow \nu\bar{\nu} \LSP)&=&21\%. & \hspace{.2in}
\end{array}
\ee
The cascade decays dominate, and unfortunately, the on-shell $W$ and
$Z$ diagrams result in a predominance of hadronic decay products,
making it more difficult to identify the primary vertex jets. However,
when there are only two quark jets, we know that these came from the
primary vertex, and of the 2333 events that pass the $\mpt$ and
$\theta_{acop}$ cuts, there are still 478 2-jet events.  Applying the
same analysis to these events as was applied to the cascade events of
region 2b in Sec.~\ref{sec:gaugino}, we find that the squark masses can
be determined to 4.8 GeV at 95\% CL.  We therefore find that despite
the predominance of cascades with hadronic decay products, we again
expect to measure the squark masses to an accuracy of about 1\%.

It is tempting to try to improve our statistics by using some of the
events with more than 2 jets. For these events, we can try to
reconstruct the on-shell $W$s and $Z$s from jet pair invariant masses
and thereby determine which quark jets are produced at the primary
vertex.  With six jets and the detector resolution of 50\% assumed
above, it is very difficult to determine with any certainty which two
of the fifteen possible quark pairs have the correct invariant masses
to be $W$ and $Z$ decay products. However, for the 996 four jet events,
the determination is much easier.  For these events, we accept only
those events where exactly one of the six quark pairs has invariant
mass within 10\% of either $m_W$ or $m_Z$. 29\% of the 4 jet events
pass this invariant mass cut, and of these, fewer than 1\% have
misidentified $W$s or $Z$s. Therefore, in 29\% of the 4$q$ and all of
the 2$q$ events we can identify the primary quarks.  Of these 779
events, 67\% are double cascade events.  We then calculate $\mmin$ for
each event and assume $\mprim=m_{\chc_1}$.  Demanding that this
hypothesis be kinematically consistent leaves 647 events, of which 80\%
are double cascades. The $\llike$ fit to the $\mmin$ distribution gives
the squark mass to 4.4 GeV at 95\% CL, a slight improvement of our
previous result.

It is important to note that backgrounds to the 4-jet events are large
and may make the 4-jet events difficult to isolate.  In particular, if
it is necessary to cut events with a pair of jets whose invariant mass
is near the $W$ or $Z$ mass to reduce the background, the analysis of
the preceding paragraph is of course not possible. In addition, if
$h^0$ is light enough to be on-shell in the cascade diagrams, $b$
quarks will dominate the decay products and one cannot anti-tag bottom
quarks.  However, by using a combination of $b$-tagging and $h^0$ mass
reconstruction one might hope to isolate the primary vertex quarks.  On
the other hand, the presence of light sleptons would make the analysis
easier, since there are then more lepton decay products and more 2-jet
events.

Measurements in region 2b with high $m_{\tilde{l}}$ and high $m_{h^0}$
will be slightly degraded for the same reasons as in 2a, namely many
cascades with few leptonic decays. One would again like to put the
large fraction of events with more than two jets to use. Unfortunately,
for these events, one will generally not be able to identify the
primary quark jets, as there are no on-shell $W$ and $Z$ decays to
reconstruct. However, it may still be possible to get information from
these events.  For example, noting that jets from cascade decays have a
smooth energy distribution concentrated at lower energies, if one plots
the distribution of 1 jet energies, one might be able to discern the
endpoints of the primary decay jet distribution above the tail from the
cascade jets. In the pessimistic high $m_{\tilde{l}}$ case we are
considering, most events will include many jets, and a realistic study
would require an accurate simulation of hadronization effects and jet
reconstruction.

\section{Additional Remarks and Conclusions}
\label{sec:conc}

In our whole analysis above, we have set $\tanb=2$.  Raising $\tanb$
has little effect on the squark branching ratios in the $\mu>0$ part of
the parameter plane, and its effect on branching ratios in $\mu<0$
regions is to make then more similar to $\mu>0$ regions. For example,
when $\tanb=20$, the only large effect is that the branching ratio to
the LSP in region 3 grows to $\approx 80\%$.  Thus, the decay patterns
are nothing new, and by exploring all regions of the plane for
$\tanb=2$, we have simultaneously roughly analyzed the case of higher
$\tanb$.

In this paper we have begun to explore the prospects for measuring
squark masses at future \epem\ colliders, within the context of the
MSSM. We have shown that, even if squarks choose complex decay
patterns, these machines do offer opportunities for making squark mass
measurements at the level of a few GeV.  Such precision measurements
would be invaluable for probing the mechanism of supersymmetry breaking
in deeper underlying theories. Interesting questions for future
research include how more realistic physics simulations would affect
this picture and how squark masses would be studied in the case that
gluinos are the dominant decay channel.

\acknowledgements
It is a pleasure to thank M.~Peskin for suggesting this topic, and we
are grateful to him and to H.~E.~Haber and T.~Barklow for many helpful
discussions.

\figure{\label{fig:sigma}
The number of squark pairs of the first two generations produced at a
500 GeV \epem\ collider with polarized beams and an integrated
luminosity of 10 $\ifb$ for each beam polarization. The four helicity
combinations plotted are $e^-_{L,R}e^+ \rightarrow
\tilde{q}_{L,R}\bar{\tilde{q}}$.}

\figure{\label{fig:plane}
The $\mum$ plane divided into regions with similar squark decay
channels, given $\tanb=2$ and $\msq=220 \gev$. The squark decays of the
various regions are described in the text. The points of parameter
space that we consider in detail in the text are marked. The condition
$m_{\sq} < M_3$ is true only above the dotted line at $M_2 = 67 \gev$.}

\figure{\label{fig:jetenergy1}
The jet energy distribution resulting from squark decays in region 1 at
the point $\mum = (-500 \gev, 300 \gev)$. Each jet produced in such
events is binned individually.  Cuts and detector resolution effects
alter the flat shape and make determination of the endpoints more
difficult. The solid (dashed) histogram represents events with $e^-_L$
($e^-_R$) polarized beams. The integrated luminosity assumed is 5
$\ifb$ per polarization, and the bin size is 4 GeV.}

\figure{\label{fig:circle}
Determination of minimum kinematically-allowed squark mass from the two
visible quark momenta and the LSP mass. The momenta label the particles
of $\sq (p_1)\rightarrow \LSP (p_3) q(p_4)$ and $\sq (p_2)\rightarrow
\LSP (p_5) q(p_6)$. The momenta of the two undetected LSPs are
constrained to lie on the circle $C$.}

\figure{\label{fig:msqmin}
The distribution of $\mmin$, the minimum allowed squark mass for a
given event, in region 1 at the point $\mum = (-500 \gev, 300 \gev)$.
The distribution for $e^-_L$ ($e^-_R$) polarized beams is given by the
solid (dashed) histogram and is sharply peaked at the actual $\sq_L$
($\sq_R$) mass of 220 (210) GeV. The integrated luminosity assumed is 5
$\ifb$ per polarization, and the bin size is 5 GeV.}

\figure{\label{fig:br}
Contours of the branching ratios
a) ${\rm BR}(\tilde{u}_L\rightarrow u\LSP)$ and
b) ${\rm BR}(\tilde{u}_R\rightarrow u\LSP)$
in percent in the $\mum$ plane. In the gaugino-like region 2,
$\tilde{u}_L$ decays predominantly via cascades, while $\tilde{u}_R$ is
seen to decay primarily to $\LSP$ even though other on-shell decays are
kinematically allowed.}

\figure{\label{fig:jetenergy2a}
Jet energy distributions at the point $\mum = (-500 \gev, 200 \gev)$ in
region 2a. Every jet produced in the Monte-Carlo squark decay events is
binned individually. The three components are a large peak of soft
quarks emanating from squarks decaying to $\chn_2$ and $\chc_1$ (long
dashes), a flat distribution from direct LSP decays (solid), and a wide
hump from hadronic $W$ and $Z$ decays (short dashes). The total
distribution is given by the dotted histogram. The analysis is
independent of the soft jets. The integrated luminosity assumed is 5
$\ifb$ per polarization, and the bin size is 4 GeV.}

\figure{\label{fig:bquarks}
The total jet energy distribution at the point $\mum = (-500 \gev, 200
\gev)$ (solid), and the bottom quark energy distribution resulting from
the top quark pair production background (dashes). In the bottom quark
spectrum, we have removed all $b$ quarks produced in top events in
which at least one $b$ quark has been successfully tagged. We assume
$m_{\rm top} = 150 \gev$ and a $b$-tagging efficiency of 80\%
\cite{JLC}.}

\figure{\label{fig:feyn}
The Feynman diagrams for three-body $\chn_2$ and $\chc_1$ decays
mediated by $h^0$, charged sleptons and sneutrinos, squarks, and $Z$
and $W$ bosons.}

\figure{\label{fig:jetenergy3}
Jet energy distributions at the point $\mum = (-100 \gev, 700 \gev)$ in
region 3. The solid (dashed) histogram represents events with $e^-_L$
($e^-_R$) polarized beams. The soft hump of secondary vertex jets is
well-separated from the primary vertex jet distribution. The integrated
luminosity assumed is 5 $\ifb$ per polarization, and the bin size is 4
GeV.}

\figure{\label{fig:1tevplane}
The $\mum$ plane divided into regions with similar squark decay
channels, given $\tanb=2$ and $\msq=400 \gev$.  The squark decays of
the various regions are as in Fig.~\ref{fig:plane} and are given in the
text. The point in region 2a that we consider in detail is marked. The
condition $m_{\sq} < M_3$ is true only above the dotted line at $M_2 =
121 \gev$.}

\end{document}